\documentclass{article}

\usepackage{arxiv}
\usepackage{amsmath}
\usepackage{graphicx}
\usepackage{cite}

\title{Optical Spectra of Plasmon--Exciton Core--Shell Nanoparticles:
 An Anisotropic Classical Model Eliminates Discrepancies with Experiments}

\author{Alexey~D.~Kondorskiy\thanks{kondorskiy@lebedev.ru} \\ \\
	P.N. Lebedev Physical Institute of the Russian Academy of Sciences,\\
	Leninskiy prosp. 53, 119991 Moscow, Russia
}

\begin{document}
\maketitle

\begin{abstract}
The optical properties of the hybrid core--shell nanostructures composed of 
a metallic core and an organic shell of molecular J-aggregates are determined 
by the electromagnetic coupling between plasmons localized at the surface of 
the metallic core and Frenkel excitons in the shell. In cases of strong and 
ultra-strong plasmon--exciton coupling, the use of the traditional isotropic 
classical oscillator model to describe the J-aggregate permittivity may lead 
to drastic discrepancies between theoretical predictions and the available 
experimental spectra of hybrid nanoparticles. We show that these discrepancies 
are not caused by limitations of the classical oscillator model itself, but by 
considering the organic shell as an optically isotropic material. By assuming 
a tangential orientation of the classical oscillators of the molecular 
J-aggregates in a shell, we obtain excellent agreement with experimental 
extinction spectra of TDBC-coated gold nanorods, which cannot be treated with 
the conventional isotropic shell model. Our results extend the understanding 
of the physical effects in the optics of metal--organic nanoparticles and 
suggest an approach for the theoretical description of such hybrid systems.
\end{abstract}

Studies of hybrid metal-organic nanoparticles have attracted considerable 
interest due to the wide range of potential applications of such structures in 
nanophotonics~\cite{Cao2018, Sun2018, Zheng2017, Wen2017, Chen2012, Li2018}. 
Nanoparticles composed of a metallic core and an organic shell of ordered J- 
and H-aggregates of molecules are of particular interest~\cite{Li2018,
Lebedev2012,Torma2015,Balci2013,DeLacy2013,Kond-Leb_OE2019,Kondorskiy2022}. 
The optical properties of such structures are determined by the electromagnetic 
near-field interaction between Frenkel excitons in the outer organic shell and 
plasmons localized at the surface of the metallic core. As the strength of this 
plasmon--excitonic (plexcitonic) coupling increases, the plasmonic band in the 
metallic core and the excitonic band in the organic shell form hybrid 
polariton states of the entire nanosystem with qualitatively new properties 
compared to those of the individual components~\cite{Torma2015}. The spectral 
properties of these polariton states can be controlled by tuning the parameters 
of the plasmonic and excitonic bands. Advances in the synthesis of plasmonic 
nanostructures allow the fabrication of metallic cores of different sizes and 
shapes with tunable plasmonic band parameters~\cite{Jiang2018}. The use of 
molecular J-aggregates of the cyanine dye in the organic shell significantly 
enhances the plexcitonic coupling in the hybrid system due to the narrow and 
intense absorption features of these aggregates~\cite{Wurthner2011,
Shapiro2018}. The size, shape and optical constants of the metallic and organic 
compounds of the nanoparticle determine different possible regimes of 
plexcitonic coupling. These regimes include weak, strong, and ultra-strong 
coupling, which have been extensively discussed in several review 
articles~\cite{Cao2018,Li2018,Torma2015,Krivenkov2019}.

Plexcitonic systems are typically simulated using classical electromagnetic 
tools such as Mie theory or numerical Maxwell equation solvers~\cite{Torma2015,
Li2018}. The dielectric functions of the materials are described by the 
Drude-Lorentz model for the metallic core and the scalar (isotropic) classical 
Lorentz oscillator model for the organic shell. When the plexcitonic coupling 
is sufficiently strong, predictions obtained with the described classical 
approach are found to disagree with experimental results~\cite{Stete2023}. 
The discrepancy is particularly noticeable in the studies of metal-organic 
structures with a shell made of the J-aggregate of the TDBC dye, since the 
value of the reduced oscillator strength of its excitonic band is large 
($f= 0.36$)~\cite{Gentile2014,Gentile2016}. The plexcitonic coupling energy 
(Rabi splitting) in the system reaches hundreds of meV. In this case, the 
above isotropic classical model predicts a third resonance between the two 
polariton bands~\cite{Lebedev2012,Antosiewicz2014}, which is associated with 
an uncoupled shell mode. However, this resonance has never been observed 
experimentally in spectra of pure metal-organic particles with a shell of 
J-aggregates (see discussion in Ref.~\cite{Stete2023}), except for samples 
containing residual uncoupled emitters~\cite{Bellessa2009}.

Regarding this problem, it is worthwhile to mention here some recent 
discussions about validity of the classical description of the nanoparticle 
excitonic subsystem. A remarkable number of works is devoted to the discussion 
of the importance of the consideration of quantum effects induced by the 
enhancement of the electromagnetic field in the vicinity of the metallic 
surface~\cite{Li2018,Torma2015,Stete2023,Gentile2016,Antosiewicz2014}. In 
Ref.~\cite{Stete2023} it was suggested that by considering the quantum effects 
one can avoid the discrepancies between the predictions of the classical model 
and the experimental data in the cases of strong and ultra-strong plexcitonic 
coupling. The heuristic quantum model introduced therein provides an 
expression for the frequency-dependent dielectric function of the shell 
material that includes the collective vacuum Rabi frequency, which depends on 
the size and shape of the structure. Using this model, the authors obtained 
excellent agreement with a series of experimental extinction spectra of 
metal-organic particles with different coupling strengths due to systematic 
size variation. To the best of our knowledge, Ref.~\cite{Stete2023} is the 
only work that explicitly addresses the problem of discrepancies between the 
predictions of the classical model and the experimental results.

In present letter, we give a completely different theoretical explanation for 
the discrepancies between the predictions of the isotropic classical model 
and the experimental results on the spectral properties of metal--organic 
particles with a shell made of the J-aggregates. We believe that the 
appearance of the uncoupled shell mode resonance in the theoretical spectra, 
which is not observed experimentally, is not a result of the limitation of the 
classical description of the optical properties of the J-aggregate as compared 
to the quantum description. This peak appears in the theoretical results under 
the assumption that the optical response of the molecular J-aggregates is 
isotropic, so that the frequency-dependent dielectric function of the organic 
shell is taken in conventional scalar form,
\begin{equation}
\varepsilon _{\text{J}}(\omega)=\varepsilon^{\infty}_{\text{J}}
+\frac{f_{\text{J}} \omega_{\text{J}}^2}{\omega _{\text{J}}^{2}-\omega ^{2}
-i\omega {\gamma_{\text{J}}}},\label{eps-J}
\end{equation}
\noindent where $\omega_{\text{J}}$ is the resonance frequency of the J-band, 
$\gamma_{\text{J}}$ is its full width, $f_{\text{J}}$ is the reduced oscillator 
strength, and $\varepsilon^{\infty}_{\text{J}}$ is the permittivity outside the 
J-band. Recently, it has been demonstrated that several experimental results 
for the optical spectra of metal-organic nanostructures~\cite{Uwada2007,
Yoshida2009a,Takeshima2020} cannot be explained in terms of the isotropic 
response model. The experimental data can be well described if the anisotropic 
and orientational effects of the outer J-aggregate shell are taken into account 
in the theoretical treatment of the optical properties of hybrid 
metal/J-aggregate and metal/spacer/J-aggregate 
nanoparticles~\cite{Kondorskiy2022}.

To confirm our statement, we perform numerical simulations of the extinction 
spectra of bare and J-aggregate coated gold spherical nanoparticles and 
nanorods considered in Ref.~\cite{Stete2023} with the dielectric function 
of an anisotropic molecular J-aggregate represented in tensor form with 
frequency dependence described by the classical Lorentz oscillator model. 
The theoretical approach used is described in Ref.~\cite{Kondorskiy2022}. 
We use an anisotropic version of Mie theory to calculate the extinction 
spectra of core-shell spherical nanoparticles. Numerical simulations of the 
extinction spectra of TDBC-coated gold nanorods are performed using the FDTD 
method implemented in the freely available software package MIT Electromagnetic 
Equation Propagation (MEEP). The final results are presented for 
orientationally averaged cross sections corresponding to naturally polarized 
light incident on the nanoparticles randomly oriented in colloidal solutions 
or matrices.

We assume axial symmetry of the optical properties of the molecular J-aggregate 
and describe the components of the dielectric function tensor of the organic 
shell in terms of $\varepsilon_{\parallel}$ and $\varepsilon_{\perp}$, which 
are the tensor components associated with the direction parallel to the 
aggregate orientation axis (\textit{longitudinal} component) and perpendicular 
to it (\textit{transverse} component), respectively. We assume that the TDBC 
J-aggregate could be described by longitudinal Lorentzian only (a general 
expression is discussed in~\cite{Kondorskiy2022}),
\begin{equation}
\varepsilon_{\parallel}\left( \omega \right) =\varepsilon^{\infty}_{\text{J}} 
+\frac{f^{\parallel}_{J}\omega^{2}_{J}}{\omega^{2}_{J}-\omega^{2}-i\omega 
\gamma_{J}}, \;  \; \; \;
\varepsilon_{\perp}\left( \omega \right) = \varepsilon^{\infty}_{\text{J}}.
\label{eps-par-perp-single}
\end{equation}
\noindent
This assumption is confirmed by experimental observations of the 
orientation-dependent optical spectra of TDBC-like aggregates, which exhibit 
anisotropic absorption with the strong sharp band with the dipole oriented 
along a single axis~\cite{Didraga2004,Pugzlys2006}.

\begin{figure*}[ht]
\centering\includegraphics[width=0.9\linewidth]{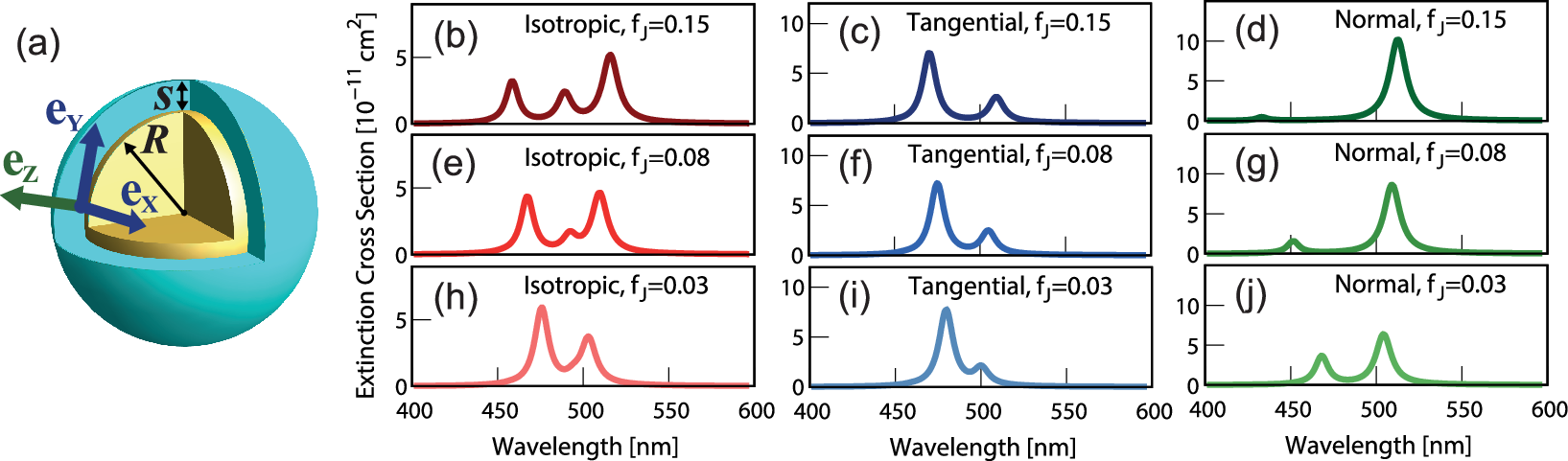}
\caption{(a) Schematic view of a Drude sphere with a dye shell under study. 
The vectors $\textbf{e}_X$, $\textbf{e}_Y$ and $\textbf{e}_Z $ show the basis 
for the permittivity tensor of the shell material: $\textbf{e}_Z $ is directed 
normal to the surface, while $\textbf{e}_X$ and $\textbf{e}_Y$ lie in the 
tangential plane. (b)--(j) Light extinction cross section of a Drude sphere 
with a dye shell with varying reduced oscillator strength, $f_{\text{J}}$, and 
different molecular arrangement in the organic shell. (b)--(d) Results are 
obtained for $f_{\text{J}} = 0.15$ and: (b) \textit{isotropic} 
(\textit{equiprobable}) J-aggregate orientation, (c) \textit{tangential} 
J-aggregate orientation, and (d) \textit{normal} J-aggregate orientation in 
the dye shell. (e)--(g) and (h)--(j), same results calculated for 
$f_{\text{J}} = 0.08$ and $f_{\text{J}} = 0.03$, respectively}
\label{fig:sphere}
\end{figure*}

\begin{figure*}[ht]
\centering\includegraphics[width=0.9\linewidth]{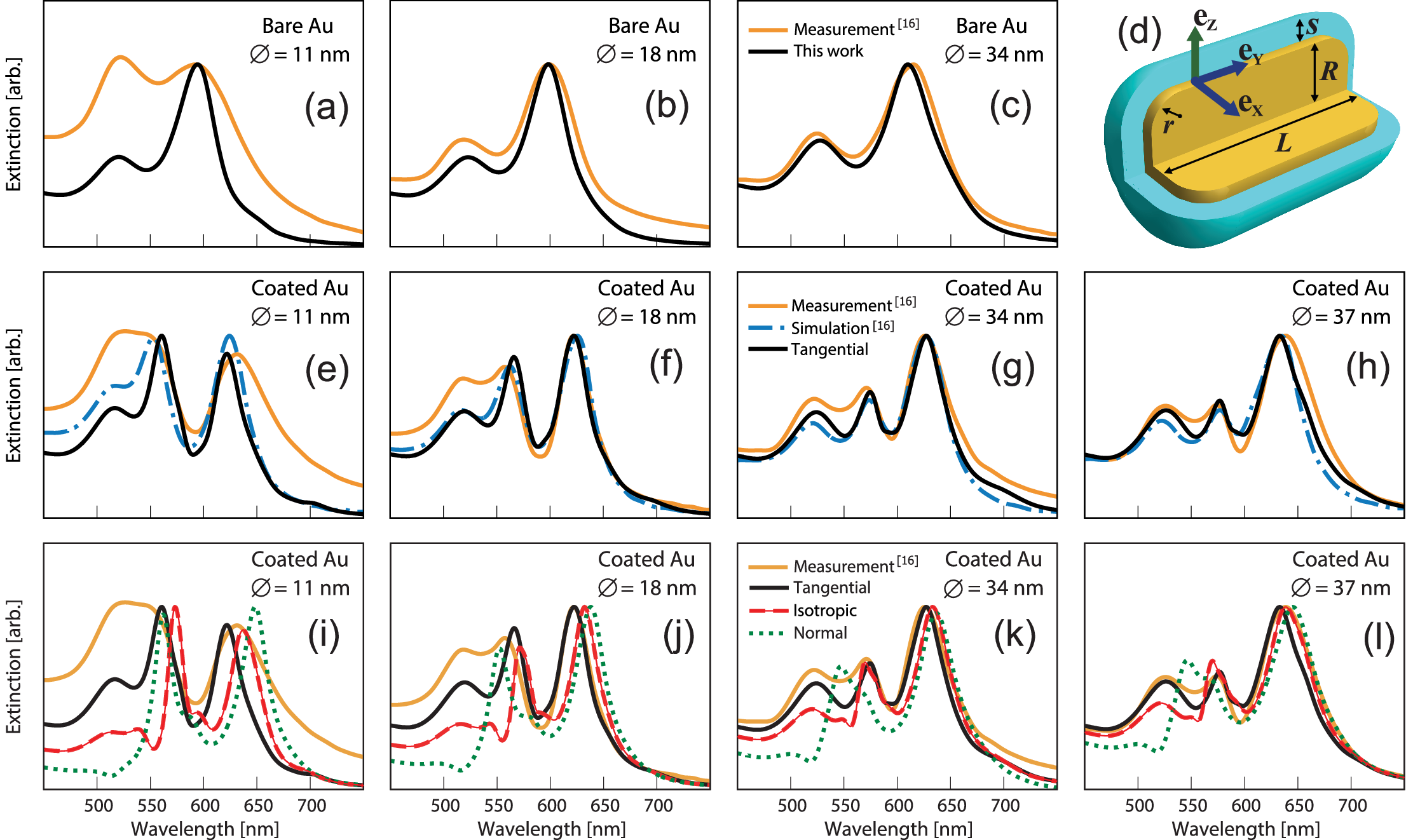}
\caption{(a)--(c) Light extinction spectra of bare nanorods of different sizes: 
(a) 11 nm, (b) 18 nm, (c) 34 nm. For the sake of brevity the results for 37 nm 
are not shown because they are similar to those for 34 nm. In these figures: 
solid orange lines -- experimental results~\cite{Stete2023}, solid black lines 
-- our calculations. (d) Schematic view of a TDBC-coated gold nanorod. (e)--(h) 
and (i)--(l) Light extinction spectra of TDBC-coated gold nanorods of different 
diameters: (e) and (i) 11 nm, (f) and (j) 18 nm, (g) and (k) 34 nm, (h) and 
(l) 37 nm. Panels (e)--(h) show comparison of experimental 
results~\cite{Stete2023} (solid orange lines) and numerical simulations 
performed with heuristic quantum model~\cite{Stete2023} (dash-dotted blue 
lines) with our results obtained with classical anisotropic model for 
\textit{tangential} J-aggregate orientation in the TDBC shell (solid black
 lines). Panels (i)--(l) show the effect of J-aggregate orientation: solid 
 orange lines -- experimental results~\cite{Stete2023}, solid black lines 
 -- \textit{tangential} orientation, thin red lines with dashes
 -- \textit{isotropic} (\textit{equiprobable}) orientation, dotted green lines 
-- \textit{normal} orientation}
\label{fig:rod}
\end{figure*}

Following Ref.~\cite{Kondorskiy2022}, we consider three cases of molecular 
arrangements in the J-aggregate shell: (I) \textit{Normal} aggregate 
orientation, when the direction of the $\varepsilon_{\parallel}$ component is 
parallel to the normal direction to a nanoparticle surface; 
(II) \textit{Tangential} aggregate orientation, in which the direction of the 
$\varepsilon_{\parallel}$ component is perpendicular to the normal direction, 
and we assume the equiprobable orientation of the aggregate axis along the 
tangential plane of a nanoparticle; (III) \textit{Equiprobable} orientation of 
the aggregate axis in space. We describe these cases with a tensor oriented 
along the $\mathbf{e}_{X}$, $\mathbf{e}_{Y}$ and $\mathbf{e}_{Z}$ axes 
(see Fig.~\ref{fig:sphere}a and Fig.~\ref{fig:rod}d). For the case of 
Eq.~(\ref{eps-par-perp-single}) the general tensor expressions are simplified 
and take the form of
\begin{equation}
\widehat{\varepsilon}\left( \omega \right) =
\varepsilon^{\infty}_{\text{J}}\widehat{I} +\frac{\omega^{2}_{J}}{\omega^{2}_{J}
-\omega^{2}-i\omega \gamma_{J}}\widehat{f}, \label{eps-tensor}
\end{equation}
\noindent
where $\widehat{I}$ is the unit tensor, and the reduced oscillator strength 
tensor, $\widehat{f}$, can be written for normal, tangential and equiprobable 
orientations, respectively, as 

\begin{equation}
\widehat{f}_{\text{norm}}=\left(\begin{array}{ccc} 0 & 0 & 0 \\ 0 & 0 & 0 
\\ 0 & 0 & f^{\parallel}_{J} \end{array}\right), \label{eq:tensor-norm}
\end{equation}
\begin{equation}
\widehat{f}_{\text{tang}}=\left(\begin{array}{ccc} 
\frac{1}{2}f^{\parallel}_{J} & 0 & 0 \\ 0 & \frac{1}{2}f^{\parallel}_{J} 
& 0 \\ 0 & 0 & 0 \end{array} \right), \label{eq:tensor-tang}
\end{equation}
\begin{equation}
\widehat{f}_{\text{epo}}=\left(\begin{array}{ccc} 
\frac{1}{3}f^{\parallel}_{J} & 0 & 0 \\ 0 & \frac{1}{3}f^{\parallel}_{J} 
& 0 \\ 0 & 0 & \frac{1}{3}f^{\parallel}_{J} \end{array} \right). 
\label{eq:tensor-epo}
\end{equation}

The case of equiprobable orientation (\ref{eq:tensor-epo}) is equivalent to the 
isotropic model (\ref{eps-J}) with the oscillator strength of the single 
longitudinal band being three times larger than the oscillator strength of the 
isotropic model. To compare the results obtained with isotropic and anisotropic 
models, we take a certain value of $f_{J}$ and assume that it can be estimated 
from the situation of the equiprobable orientations that take place in 
solution of the molecular aggregates. The results for normal and tangential 
orientations are obtained with the 
Eqs.~(\ref{eq:tensor-norm})--(\ref{eq:tensor-tang}) with 
$f^{\parallel}_{J} = 3 f_{J}$. Because of this relation, we have a single 
value of $f_{J}$ to refer to parameters of both the isotropic and anisotropic 
models used.

First, we perform calculations of the light extinction cross section for a 
model Drude sphere with a dye shell in water. Our results for this system are 
shown in Figure~\ref{fig:sphere}. The permittivity of the core is described by 
the Drude model for gold, and the parameters for the permittivity of the 
J-aggregate shell are $\varepsilon^{\infty}_{\text{J}}=1.7$, 
$\omega_{\text{J}}=2.5$ eV, (wavelength $\lambda = 495$ nm) 
$\gamma_{\text{J}}=0.05$ eV. The geometrical parameters are $R=12.5$ nm and 
$s=3$ nm (see Fig.~\ref{fig:sphere}a). The results of calculations performed 
with the \textit{isotropic} classical description of the dye permittivity 
clearly show the growth of the uncoupled shell band around $\lambda = 492$ nm 
with increasing reduced oscillator strength, $f_{J}$. However, there is no 
such band in the results obtained with the anisotropic shell model for both 
the \textit{tangential} and \textit{normal} J-aggregate orientations in 
the organic shell.

Second, we perform calculations of the extinction spectra of bare and 
TDBC-coated gold nanorods of different sizes as measured and simulated in 
Ref.~\cite{Stete2023}. Our results are shown in Figure~\ref{fig:rod}. The 
notation of the sizes of the nanorods studied corresponds to the transverse 
diameters of 11, 18, 34, and 37 nm as measured in Ref.~\cite{Stete2023}. We 
use the following actual sizes of the nanorods: for a transverse diameter of 
11 nm, $2R=11.5$ nm, $L=19.5$ nm, $r=5$ nm; for diameter 18 nm $2R=19.5$ nm, 
$L=32$ nm, $r=7$ nm; for diameter 34 nm $2R=36$ nm, $L=58$ nm, $r=14$ nm; 
for diameter 37 nm $2R=41$ nm, $L=66$ nm, $r=15$ nm. The permittivity of gold 
was taken from Ref.~\cite{Olmon2012}. For bare nanorods, a 1 nm thick citrate 
capping layer with permittivity of 2.37 is included (see Ref.~\cite{Park2014}). 
The TDBC shell thickness is $s=3$ nm. For the TDBC J-aggregate we use the 
parameters of the Lorentz oscillator model, which are in agreement with those 
previously reported~\cite{Stete2023,Bellessa2009,Gentile2014}: 
$\varepsilon^{\infty}_{\text{J}}=2.3$, $\omega_{\text{J}}=2.12$ eV, 
($\lambda = 585$ nm) $\gamma_{\text{J}}=0.048$ eV. $f_{\text{J}} = 0.22$. 
We use the same parameters of the dielectric function of the TDBC J-aggregate 
for all nanorod sizes in our calculations. The permittivity of the environment 
was set to 1.96.

Figure~\ref{fig:rod} shows our results for the extinction spectra of bare and 
TDBC-coated gold nanorods. It is seen that the results obtained for the 
tangential J-aggregate orientation in the organic shell of the TDBC-coated 
gold nanorods are in good agreement with the experimental 
data~\cite{Stete2023}. Our calculations for the cases of isotropic 
(equiprobable) and normal J-aggregate orientations do not show a sufficient 
quality of agreement with the experimental data for all nanorod sizes. 
Therefore, we conclude that the J-aggregates of the TDBC dye are assembled 
on the metallic surface so that their axis is directed along the surface. Note 
that the uncoupled shell resonance appears in the spectra calculated for 
isotropic J-aggregate orientation around $\lambda = 585$ nm 
(see Figs.~\ref{fig:rod}i -- \ref{fig:rod}j).

The predictions of the anisotropic classical model have a similar quality of 
agreement with experimental data as those of the heuristic quantum 
model~\cite{Stete2023} (see Fig.~\ref{fig:rod}). However, our model is much 
simpler because it does not use quantum effects. Also, this classical model 
does not require \textit{ad hoc} modification of the spectroscopic parameters 
of the organic material due to its assembly on the metal surface, while the 
heuristic quantum model does. We emphasize that our theoretical results are 
obtained with the same set of optical parameters of the excitonic subsystem 
for all nanostructure sizes.

In conclusion, we report the results of theoretical studies of the optical 
properties of hybrid metal-organic nanoparticles with a metallic core and an 
organic shell made of molecular J-aggregates. We address the problem of drastic 
discrepancies between the theoretical predictions of the traditional isotropic 
classical oscillator model and some of the available experimental spectra, 
which appear in such systems in cases of strong and ultra-strong 
plasmon--exciton coupling. These discrepancies are manifested in the difference 
in the number of spectral bands predicted by theory and observed 
experimentally. The traditional classical model predicts a third resonance 
between the two polariton bands, which is not observed experimentally. We show 
that the observed discrepancies are not due to the limitations of the classical 
oscillator model, but to the treatment of the organic shell as optically 
isotropic medium.

Since optical anisotropy is an essential feature of the molecular 
J-aggregates, we have studied the effects of the orientation of the 
molecular J-aggregates in the organic shell on the optical spectra of 
the hybrid nanostructure. We describe anisotropic molecular J-aggregates 
with a dielectric function represented in tensor form with a frequency 
dependence described by the classical Lorentz oscillator model. We show that 
in the case of strong and ultra-strong plasmon--exciton coupling, the 
extinction spectra of the hybrid nanoparticle differ significantly for 
isotropic, tangential and normal orientations of the molecular J-aggregate 
in the organic shell. While the third peak is observed in the case of the 
isotropic orientation, it is not observed in either the case of the normal 
orientation or the tangential orientation. Furthermore, we show that the 
experimental extinction spectra of TDBC-coated gold nanorods~\cite{Stete2023} 
can be excellently reproduced if one assumes a tangential orientation of the 
molecular J-aggregates in shell when describing the permittivity of the 
J-aggregate with the use of the classical oscillator model. However, this 
system cannot be treated with the conventional isotropic shell model.

Our results clearly demonstrate the importance of incorporating the 
anisotropic optical properties of J-aggregates into the theoretical 
description of the optical properties of hybrid metal-organic core-shell 
nanostructures.

The author is grateful to V. S. Lebedev and A.A. Narits for the valuable 
discussions. This work was supported by the Russian Science Foundation under 
grant No. 19-79-30086.

\end{document}